\documentstyle[prb,preprint,aps,epsfig]{revtex}

\tightenlines
\begin{document}
\draft
\title{Friction force on a vortex due to the scattering of superfluid
 excitations in helium II }

\author{H. M. Cataldo and D. M. Jezek}

\address{Departamento de F\'{\i}sica, Facultad de Ciencias Exactas y Naturales, \\
Universidad de Buenos Aires, RA-1428 Buenos Aires, Argentina\\
and Consejo Nacional de Investigaciones Cient\'{\i}ficas y
T\'ecnicas, Argentina}

\date{\today}
\maketitle

\begin{abstract}
The longitudinal friction acting on a vortex line in superfluid $^4$He is investigated 
within a simple model based on the analogy 
between such vortex dynamics and that of the quantal
Brownian motion of a charged point particle in a uniform magnetic field. The scattering of
superfluid quasiparticle excitations by the vortex stems from a translationally invariant 
interaction potential which, expanded to first order in the vortex velocity operator, gives 
rise to vortex transitions between nearest Landau levels. The corresponding friction coefficient
is shown to be, in the limit of elastic scattering (vanishing cyclotron frequency),
equivalent to that arising from the Iordanskii formula. Proposing a simple functional form 
for the scattering amplitude, with only one adjustable parameter
whose value is set in order to get agreement to
the Iordanskii result for phonons, 
an excellent agreement is also found with the values derived from 
experimental data up to temperatures about 1.5 K. Finite values of the cyclotron 
frequency arising from recent theories are shown to yield similar results.
The incidence of vortex-induced quasiparticle transitions on the friction process is estimated
to be, in the roton dominated regime, about 50 \% of the value of the friction coefficient,
$\sim$8 \% of which corresponds to roton-phonon transitions and $\sim$42 \% to roton 
$R^+\leftrightarrow R^-$ ones.
\end{abstract}

\section{Introduction}
\label{sec1}

The dynamics of quantized vortices in helium II constitutes one of the most remarkable
topics in the study of superfluidity, but only at absolute zero such dynamics is well
understood. At finite temperatures, on the other hand, the motion of vortices is subjected
to frictional dissipation arising from the interaction with thermally excited quasiparticles
and, while there exists an acceptable empirical knowledge of this phenomenon, the microscopic
understanding of it leaves much to be desired, as we shall briefly survey in the following.

At zero temperature the force per unit length acting on a vortex line parallel to the
$z$-axis, moving with a velocity ${\bf v}$, is the so-called Magnus force,\cite{don}
\begin{equation}
{\bf F}_M=\rho_s\kappa\,\hat{{\bf z}}\times({\bf v}-{\bf v}_s).
\label{M}
\end{equation}
Here  $\rho_s$ is the superfluid mass density,
$\kappa=h/m$ denotes one quantum of circulation
($m$ being the mass of a $^4$He atom
and $h$ Planck's constant),  
and ${\bf v}_s$ takes into account an eventual uniform superfluid velocity
far from the line.

According to the two-fluid model, at nonvanishing temperatures the vortex line 
 moves through a gas of excitations, phonons and
rotons, and the friction owing to collisions with such quasiparticles gives rise to a drag
force ${\bf F}_d$ that must be added to the Magnus force,\cite{don} 
\begin{equation}
{\bf F}_d=-D\,\hat{{\bf z}}\times[\hat{{\bf z}}\times({\bf v}_n-{\bf v})]
+D'\,\hat{{\bf z}}\times({\bf v}_n-{\bf v}),
\label{FD}
\end{equation}
where ${\bf v}_n$ denotes the normal fluid velocity. It is important to notice that for
temperatures above about 1 K, the dragging of normal fluid by the vortex becomes appreciable, 
and so the normal fluid velocity at the core of the vortex will differ from that measured far 
from the core.\cite{h-v}
Consequently, the longitudinal (i.e., parallel to the vortex
velocity) and transverse
friction coefficients, $D$ and $D'$, will take different values depending on whether the 
local or bulk normal fluid velocity is considered in formula (\ref{FD}). In the first case,
we refer to such coefficients as the {\em microscopic} ones, whereas in the second we refer 
to them as the {\em phenomenological} ones.\cite{bar} 
Different theoretical models have been proposed to account for the transverse 
coefficient\cite{varios} and a considerable controversy has arisen between them, since
 the experimental data so far obtained has not been able to provide 
enough hints to prove their validity.
On the other hand, the state of the art is much more satisfactory
 for the
longitudinal coefficient. In the lowest temperature range
we begin by mentioning the pioneering study of Rayfield and 
Reif,\cite{reif} who measured  such coefficient for 0.28 K $<T<$ 0.7 K
 and deduced the value of the roton scattering cross section.
The phonon scattering contribution, by contrast, could not be determined because of the
scant information arising from the low temperature measurements, where phonons and
$^3$He impurities strongly compete in the vortex drag force mechanism.
Unfortunately, such a situation persists, since  no conclusive 
 results on this matter have been reported so far. 
For temperatures above 0.6 K, however, the phonon and $^3$He contributions
should be negligible and the drag force is then ascribed to roton scattering.
In such a case it was later found that
the validity of the Rayfield-Reif results can be extended to higher temperatures up to 1.3 K
approximately,
with some evidence of a slow increase of the roton scattering cross section value with
temperature.\cite{dijk}

All the above results for temperatures below 1.3 K are based on experiments which measure
the drag on large vortex rings.\cite{reif,dijk}
On the other hand, for higher temperatures up to the lambda
transition, the value of the friction coefficients is deduced from experiments involving
attenuation of second sound in rotating containers that yield measurements of mutual 
friction on rectilinear vortices.\cite{bar}

The most reliable theoretical study of the phonon drag force
seems to be the 1965 Iordanskii's
calculation\cite{ior} which confirmed the $T^5$ temperature dependence of the longitudinal
coefficient found previously by Pitaevskii,\cite{pita} except for a correction in the
value of the proportionality constant. 
Actually the Iordanskii formula can in principle be applied to the
whole dispersion range, i.e. to any superfluid excitation. His study was based on a model
of a weakly interacting Bose gas leading to an expression for the longitudinal coefficient
which depends on the quasiparticle-vortex scattering amplitude.
While the phonon-vortex 
scattering amplitude was considered in detail by Iordanskii, in the roton
case he made use of a quasiclassical approximation that yielded poor quantitative agreement
with the longitudinal friction figures deduced from experiments. 
It is worth mentioning also, that the Iordanskii formula
has recently been used to model the dissipative dynamics of vortices in trapped Bose-condensed
gases.\cite{fedi}

Within the last decade, there has been several theoretical approaches to the calculation
of the longitudinal force, ranging from semiempirical treatments based on numerical 
simulations of roton-vortex interactions,\cite{sam} to more formal results studying the
equivalence of adiabatic perturbation and kinetic approaches.\cite{t-t} Some other studies
on vortex dynamics in two dimensions were based on the analogy between a quantized vortex
in a superfluid and an electron in a uniform magnetic field.\cite{demi,arovas}
Such an analogy stems from the fact that the Lorentz and
Magnus forces are both transverse forces whose magnitude is proportional to the velocity.
Thus the cyclotron frequency of such vortex motion should be inversely proportional to a
vortex effective mass whose value has also been a topic of debate.\cite{duan,niu} Moreover,
only very recently a quantum theory starting from first principles, which supports the 
existence of such cyclotron motion, has been proposed showing that there are rotational states
of a quantized vortex which form  highly degenerate energy levels similar to the Landau
levels in the integer quantum Hall effect.\cite{tang}

The idea of a vortex dynamics ruled
by a cyclotron motion is certainly a very simple and appealing one. This led us to begin the
study of
a quantal Brownian motion model\cite{ca1}
by which the vortex is regarded as a quantum particle coupled
to the heat reservoir composed by the quasiparticle (qp) excitations of the superfluid. 
 Thus, it has been shown\cite{ca2} that the only
linear coupling on the vortex degrees of freedom, position and momentum,
 that leads to a dynamics 
consistent with the drag force (\ref{FD}), corresponds to the vortex
velocity. With respect to
 the qp dependence of the coupling, we have recently\cite{ca3} begun to
study an interaction Hamiltonian based on the simplest scattering events,  consisting in
 qp-vortex collisions which yield an outgoing qp, making the vortex to raise or lower one
Landau level, the latter being equivalent to the  above mentioned vortex velocity part of the 
coupling. In such a preliminary study we focused on temperatures below 0.4 K at which 
only phonons are thermally excited.\cite{he3}
In the present paper we shall concentrate our attention on the longitudinal friction 
coefficient, 
extending  our previous approach to include
the whole dispersion curve, up to temperatures at which the simple non-interacting qp gas 
picture is expected to break down ($T <$ 1.7 K).

 In the next section, starting from a generic
translationally invariant vortex-qp interaction potential, we first extract a suitable
approximation to first order in the vortex velocity. Then, making use of previous results,
the longitudinal friction coefficient is calculated, showing that it
turns out to be equivalent to that arising from
the Iordanskii formula, in the limit of a vanishing cyclotron frequency. 
A very simple functional form
for the scattering amplitude, which is proposed for the whole qp dispersion range, leads 
in Section \ref{sec3} to a thorough study of  the friction coefficient, 
together with a comparison to the results derived from experiments and an analysis of
  the incidence of vortex-induced qp transitions
of different kinds on the friction process. Finally, a summary and the main conclusions of this
study are gathered in Section \ref{sec4}.

\section{The model}
\label{sec2}
To begin with, we note that for a vanishing superfluid velocity,
${\bf v}_s=0$, the Magnus force, Eq. (\ref{M}), takes the form of
a Lorentz force acting upon an electron in a uniform magnetic field. Then classically one has
a cyclotron motion dynamics and, quantum mechanically, an energy spectrum of equally spaced 
Landau levels,\cite{cohen} both pictures ruled by a cyclotron frequency:
\begin{equation}
\Omega=\rho_s\kappa\frac{L}{M},
\label{om}
\end{equation}
where $L$ and $M$ respectively denote the vortex length and effective mass. Therefore, the 
Hamiltonian from which the Magnus force derives can be written as,\cite{cohen}
\begin{equation}
H_M=\frac{1}{2}M{\bf v}^2=\hbar\Omega\left(a^\dagger a+\frac{1}{2}\right),
\label{hm}
\end{equation}
where
\begin{equation}
\label{5}
{\bf v}={\bf p}/M+\frac{\Omega}{2}\hat{{\bf z}}\times {\bf r}
\end{equation}
gives the vortex velocity operator in terms of position ${\bf r}$ and momentum ${\bf p}$
operators, and
\begin{mathletters}
\begin{eqnarray}
v_x & = & i\sqrt{\frac{\hbar\Omega }{2M}}(a^\dagger - a) \\
v_y & = & \sqrt{\frac{\hbar\Omega }{2M}}(a^\dagger + a)
\end{eqnarray}
\label{vcomp}
\end{mathletters}
\noindent
gives the velocity components in terms of creation ($a^\dagger$) and annihilation ($a$)
 operators
of right circular quanta. Note that a possible temperature
dependence of the cyclotron frequency, arising from the factor $\rho_s$ in Eq. (\ref{om}),
can be avoided by working at low enough temperatures. 
For instance, the superfluid density $\rho_s$
changes less than 10 \% approximately for temperatures below 1.5 K.

The Hamiltonian of the whole system (vortex plus qp gas) is formed by a 
vortex term Eq. (\ref{hm}), plus a non-interacting qp term, plus an interaction
potential term which should take into account the vortex-qp scattering processes. 
The latter
may be modeled through the following translationally invariant form:
\begin{equation}
\label{ip}
\sum_{ {\bf k} , {\bf q} }  \, \, \delta_{k_zq_z}\,\Lambda_{{\bf k}  {\bf q}} 
\exp[-i{\bf r}\cdot({\bf k}-{\bf q})]
a_{{\bf k}}^\dagger \,  a_{ {\bf q}},
\end{equation}
where $a_{{\bf k}}^\dagger$ ($a_{{\bf k}}$) denotes the
 creation (annihilation) operator of a qp of
momentum $\hbar{\bf k}$, $\Lambda_{{\bf k}  {\bf q}}$ denotes a coupling parameter
to be determined, and the Kronecker delta factor assumes that there is no momentum 
transfer along the vortex line.\cite{bar}
 The form (\ref{ip}), however, is difficult to deal with due to the
nonlinear dependence on the vortex coordinate. Such a drawback can be overcome by making
the replacement,
\[{\bf r}\rightarrow {\bf r}+\frac{2}{M\Omega}{\bf p}\times\hat{{\bf z}}=
\frac{2}{\Omega}{\bf v}\times\hat{{\bf z}}\]
in the argument of the exponential in Eq. (\ref{ip}). This modified potential preserves
 translational invariance and, on the other hand, it makes possible
a linear approximation for the usually
small values of the vortex velocity, leading to an 
interaction potential of the form, 
\begin{mathletters}
\begin{equation}
\label{8a}
V=-{\bf B}\cdot{\bf v}
\end{equation}
with,
\begin{equation}
{\bf B}=\frac{2i}{\Omega}\hat{{\bf z}}\times
\sum_{ {\bf k} , {\bf q} }  \, \, \delta_{k_zq_z}
\Lambda_{{\bf k}  {\bf q}} 
({\bf k}-{\bf q}) a_{{\bf k}}^\dagger \,  a_{ {\bf q}}.
\end{equation}
\label{int}
\end{mathletters}
It is not difficult to have an estimation of the upper value of the vortex 
velocity that allows a linear approximation for 
$\exp[-i\frac{2}{\Omega}{\bf v}\times\hat{{\bf z}}\cdot ({\bf k}-{\bf q})]$.
Taking $|{\bf k}-{\bf q}|$ of the order of the roton momentum, and $\Omega\gtrsim$ 0.01
ps$^{-1}$ (see Refs. 14 and 16), the argument of the exponential turns
out to be less than $v$/(25 cm/s). Note that such an 
approximation yields a zero-th order term, 
$\sum_{ {\bf k} , {\bf q} }  \, \, \delta_{k_zq_z}\,\Lambda_{{\bf k}  {\bf q}}\,\, 
a_{{\bf k}}^\dagger \,  a_{ {\bf q}}$ (cf. Eq. (\ref{ip})), that should be added to the 
non-interacting qp Hamiltonian $\sum_{ {\bf k}}\,\hbar\omega_k \,\,
a_{{\bf k}}^\dagger \,  a_{ {\bf k}}$.
However, it will be shown later that the coupling parameter
$\Lambda_{{\bf k}  {\bf q}}$ turns out to be by far much smaller than the qp energies.

Here it is worthwhile recalling that an interaction potential of the form (\ref{8a})
(i.e., proportional to the vortex velocity, Eq. (\ref{5})), was found to be the only linear 
combination of vortex position and momentum operators leading to a dynamics in agreement with
a drag force of the form (\ref{FD}).\cite{ca2}

It is instructive to identify the scattering processes embodied in the interaction potential
(\ref{int}); in fact if we replace the velocity from Eqs. (\ref{vcomp}), we find that the 
interaction consists of terms of the form $a^\dagger a_{{\bf k}}^\dagger a_{{\bf q}}$
and $a\, a_{{\bf k}}^\dagger a_{{\bf q}}$, i.e. vortex-qp collisions that make the 
vortex to raise or lower one Landau level.

The quantum mechanical problem posed by the vortex Hamiltonian (\ref{hm}), plus a 
non-interacting qp Hamiltonian, plus the interaction (\ref{int}), complemented by a thermal
equilibrium assumption for the qp gas, can be solved for the expectation value of the
vortex position operator.\cite{ca3}
In fact, making use of  the Markov approximation to 
second order in the coupling parameter, which is assumed to depend only on
the modulus of the qp momentum
 ($\Lambda_{{\bf k}  {\bf q}}=\Lambda(k,q)$),
we obtain the following expression for the microscopic friction coefficient,
\begin{equation}
D_\Omega = \frac{2\pi}{\hbar\Omega L}
\sum_{ {\bf k} , {\bf q} }  \, \, \delta_{k_zq_z}
|\Lambda(q,k)|^2
({\bf k}-{\bf q})^2 [n(\omega_k)-n(\omega_q)]
\delta(\omega_q-\omega_k-\Omega),
\label{9}
\end{equation}
where $n(\omega)=[\exp(\hbar\omega/k_BT)-1]^{-1}$ 
denotes the Bose occupation number for the 
qp excitations. The argument of the above Dirac delta shows the energy conservation
rule corresponding to the scattering processes, e.g., a qp of energy $\hbar\omega_k$
combines with a vortex energy quantum, $\hbar\Omega$, to form a qp of energy 
$\hbar\omega_q=\hbar\omega_k+\hbar\Omega$. Turning in Eq. (\ref{9})
to the continuum limit according to
$\sum_{ {\bf k} , {\bf q} }  \, \, \delta_{k_zq_z}\rightarrow
\frac{L\,A^2}{(2\pi)^5}\int d^3{\bf k}
\int d^3{\bf q}\,\,\, \delta(k_z-q_z)$, we have,
\begin{equation}
D_\Omega =\frac{A^2}{(2\pi)^4\hbar\Omega}\int d^3{\bf k}
\int d^3{\bf q}\,\,\, \delta(k_z-q_z)\,
|\Lambda(q,k)|^2
({\bf k}-{\bf q})^2
[n(\omega_k)-n(\omega_q)]\,\,
\delta(\omega_q-\omega_k-\Omega),
\label{10}
\end{equation}
where $A$ denotes the area of the system in the $x-y$ plane.

We note that within this simple model we 
are unable to take into account the dragging of normal fluid by the vortex. 
The qp gas was in fact assumed to remain at rest, which should be a good approximation
at least for temperatures up to about 1 K.\cite{bar}

The expression (\ref{10}), which is the main result of this paper, will be compared to the 
Iordanskii formula valid for small values of the vortex velocity relative to the qp 
gas,\cite{ior}
\begin{equation}
D_I=-\int \,\frac{\partial n}{\partial\omega_k}\,\frac{k_x}{\hbar}\lim_{\rho\rightarrow
\infty}\left\{
\int \,\frac{\delta(\omega_k-\omega_q)}{\delta(k_z-q_z)}\,(k_x-q_x)\,
|f_\rho({\bf q},{\bf k})|^2\,\frac{d^3{\bf q}}{(2\pi)^3}\right\}\frac{d^3{\bf k}}{(2\pi)^3},
\label{11}
\end{equation}
where without loss of generality we have assumed a vortex velocity along the 
$x$-axis. The scattering amplitude $f_\rho({\bf q},{\bf k})$ corresponds to a vortex-qp
interaction potential which is cut off to zero at a distance $\rho$ from the vortex center.
Note that the Dirac delta in the denominator of the integrand only makes
sense if $f_\rho({\bf q},{\bf k})$ is proportional to
$\delta(k_z-q_z)$.
At low temperatures only phonon-vortex scattering is relevant, and  the limiting
process  and the integral over the wave vector
${\bf q}$ in Eq. (\ref{11}) commute\cite{ior}, so we have,
\begin{equation}
D_I =
-\frac{1}{(2\pi)^6\hbar}
\int d^3{\bf k}\,\int d^3{\bf q}\,\,\frac{\partial n}{\partial\omega_k}
\,\,\delta(\omega_k-\omega_q)\,\,k_x(k_x-q_x)\,\,
 \delta(k_z-q_z)\,
|\tilde{f}({\bf q},{\bf k})|^2.
\label{12p}
\end{equation}
The phonon-vortex scattering amplitude
 $f({\bf q},{\bf k})\equiv\tilde{f}({\bf q},{\bf k})\,\delta(k_z-q_z)$ 
was extracted by Iordanskii\cite{ior}, and its replacement in Eq. (\ref{12p}) yields,
\begin{equation}
D_I=\frac{76}{7}\frac{\zeta(5)(k_BT)^5}{\hbar^2c_s^7m^2},
\label{12}
\end{equation}
where $c_s$ denotes the sound velocity and $\zeta(n)$ the Riemann zeta function.

Given the difficulties in the calculation of a reliable expression for the scattering amplitude
outside the phonon range, we shall assume in what follows that a simplified form like 
Eq. (\ref{12p}) stands for the whole dispersion range.
To compare Eq. (\ref{12p}) with our formula (\ref{10}),  we first note that taking 
the limit
$\Omega\rightarrow 0$ in Eq. (\ref{10}),
\begin{equation}
D_0=\lim_{\Omega\rightarrow 0}D_\Omega =
-\frac{A^2}{(2\pi)^4\hbar}
\int d^3{\bf k}\,\int d^3{\bf q}\,
\,\frac{\partial n}{\partial\omega_k}
\,\,\delta(\omega_k-\omega_q)\,\,({\bf k}-{\bf q})^2\,\,
 \delta(k_z-q_z)\,
|\Lambda(q,k)|^2,
\label{13}
\end{equation}
an expression very similar  to (\ref{12p}) is obtained. Actually, the factor 
$\delta(\omega_k-\omega_q)$ in Eqs. (\ref{11}) and (\ref{12p})
corresponds to the common assumption of
elastic scattering of qp excitations by a vortex line.\cite{don,bar,reif,ior,sam}
On the other hand, a vanishing cyclotron frequency is clearly inadmissible within our
model, but if we assume a gap between Landau levels that scales down with the vortex 
size,\cite{arovas,duan,tang} 
such a limit may be regarded as a good approximation for a macroscopic
system. In fact, Duan\cite{duan} based on arguments of gauge-symmetry breaking, proposed for
a vortex line in superfluid $^4$He an effective mass per unit length logarithmically
 divergent with the sample size, which for sizes of order $\sim 1$ cm yields
$\Omega\simeq$ 0.15 ps$^{-1}$ (Eq. (\ref{om})).
Arovas and Freire\cite{arovas} exploiting the analogy to (2+1)-dimensional electrodynamics,
showed that such a divergence is actually cut off at nonzero frequencies, with the wavelength
of sound at that frequencies serving as a length scale for the cut-off.
Finally, Tang\cite{tang} in a quite different approach showed 
 that in two dimensions the level spacing of a 
vortex increases with the energy and scales down logarithmically with the vortex size.
According to his estimation of the gap between the lowest levels, we have
$\Omega\simeq$ 0.011 ps$^{-1}$ for a vortex in a $^4$He superfluid film of size $\sim 1$ cm.
In the next section we shall compare $D_0$ to $D_\Omega$, with $\Omega$ given by the above
figures.

Given that the integrand of Eq. (\ref{13}) is isotropic in the $k_x$-$k_y$ and $q_x$-$q_y$
planes, the factor
$({\bf k}-{\bf q})^2$ can be replaced by $2k_x(k_x-q_x)+2q_x(q_x-k_x)$ and, taking into
account the invariance of the rest of the integrand under the replacement
${\bf k}\leftrightarrow{\bf q}$, both terms can in turn be replaced by $4k_x(k_x-q_x)$.
This means that if we identify 
\begin{equation}
|\tilde{f}({\bf q},{\bf k})|^2
\equiv (4\pi A)^2|\Lambda(q,k)|^2,
\label{fqk}
\end{equation}
$D_I$ (expression (\ref{12p}))  can be shown to be 
equivalent to $D_0$. 

Most of the integration in Eq. (\ref{10}) can be analytically performed in
spherical coordinates, leading to the 
following one-dimensional integral:
\begin{mathletters}
\begin{equation}
D_\Omega=\left(\frac{A}{2\pi}\right)^2\frac{2}{\hbar\Omega}
\,\int_0^\infty dk\,k\,[n(\omega_k)-n(\omega_k+\Omega)]\,
\sum_j \frac{|\Lambda(q_j,k)|^2}{|\omega'_{q_j}|} \,
G(q_j,k)
\label{16a}
\end{equation}
\begin{equation}
 G(q,k)=   \left\{
\begin{array}{r}
              kq(q^2+k^2/3)\,\,\,\,\,\,(k\leq q) \\
              q^2(k^2+q^2/3)\,\,\,\,\,\,(q\leq k)
            \end{array}
\right.
\label{16b}
\end{equation}
\label{D0}
\end{mathletters}
where $\omega'_q=\partial \omega_q/\partial q$ denotes the group velocity arising from the
identity
\begin{equation}
\delta(\omega_q-\omega_k-\Omega)=\sum_j\frac{\delta(q-q_j)}
{|\omega'_{q}|}\,,
\label{delta}
\end{equation}
and the summation runs over the values $q_j$ fulfilling $\omega_{q_j}=\omega_k+\Omega$.
Note, however,  that the group velocity in the denominator of Eq. (\ref{16a}) vanishes at 
the roton
and maxon points, and a simple inspection of the integrand of such an equation shows that the
only healing to such divergences can stem from the coupling parameter $\Lambda(q,k)$.
The simplest form that meets such a requirement is
\begin{equation}
|\Lambda(q,k)|^2=\alpha\,|\omega'_q||\omega'_k|\,,
\label{15}
\end{equation}
where the value of the scale parameter $\alpha=(19/560)
(2\pi\hbar^2/m\,A\,c_s)^2$ was set to reproduce the 
result (\ref{12}) from Eq. (\ref{D0}) in the limit of a vanishing cyclotron frequency
and the phonon dominated regime ($T\lesssim 0.4$ K).
Actually, any power greater than unity of the group velocities in Eq. (\ref{15})
could avoid the 
divergencies, but we have found that the first power yields the best fit to the
experimental results on $D$. Note that taking into account in Eq. (\ref{15})
a highest group velocity of the order of the sound 
velocity,\cite{dd} the maximum value of $|\Lambda(q,k)|/\hbar$ turns out to be about
$\hbar/mA\sim 10^{-16}$ ps$^{-1}$ for $A\sim$ cm$^2$,  a value which is by far much
smaller than $\Omega$ and any characteristic qp frequency (see Fig. 1). Such figures are
consistent with the assumption from which Eq. (\ref{9}) was derived, i.e., a Markov 
approximation to second order in the coupling parameter.
Finally, replacing the form (\ref{15}) in Eq. (\ref{D0}) and taking the limit $\Omega
\rightarrow 0$ we have,
\begin{equation}
              D_0
=-\frac{19}{280}\,\frac{\hbar^3}{m^2c_s^2}\int_0^\infty dk\,k|\omega'_k|\,
\frac{\partial n}{\partial\omega_k}\,
\sum_j G(q_j,k), 
\label{16}
\end{equation}
with $G(q,k)$ given in Eq. (\ref{16b}).

In the next section we shall analyze the above results. 

\section{results}
\label{sec3}
Firstly it is expedient to recall our definition of 
the wave vectors $q_j$ appearing in Eq. (\ref{16}):
given the absolute value of a qp wave vector $k$ and its corresponding energy 
$\hbar\omega_k$, the equation $\omega_q=\omega_k$ may have one, two , or three solutions in
the variable $q$, which we call $q_j$
($j=$1, 2, 3). In fact, (see Fig. 1) for $\omega_k$ below the
roton minimum there is only one solution, $q_1=k$, and the same occurs for $\omega_k$ above
the maxon value. For $\omega_k$ lying in between the roton and maxon values there are
three solutions, while for $\omega_k$ equating any of such values there are two solutions.

The physical meaning of these solutions represents the different possibilities of 
qp-vortex interaction processes, namely, {\em pure scattering events} at which the 
qp momentum is simply deflected conserving its absolute value, or {\em vortex-induced qp
transitions} changing the absolute value of momentum but conserving energy. 
For instance, a phonon of the highest part of the linear
portion of the dispersion curve (e.g., point P in Fig. 1), may undergo three different kinds
of interaction with the vortex: (i) a pure scattering event at which  the absolute 
value of momentum is conserved; 
(ii) a phonon$\rightarrow$roton $R^-$ transition at which the outgoing 
qp corresponds to a roton at the left side of the roton minimum\cite{not1} (Fig. 1);
(iii) a phonon$\rightarrow$ roton $R^+$ transition. Similarly, an $R^\pm$ roton could
be simply scattered conserving the absolute value of its momentum, or could undergo a 
vortex-induced transition to a phonon or to an $R^\mp$ roton.

The above picture corresponds to the limit of elastic scattering of qp excitations, where 
the value of the vortex energy quantum (cyclotron frequency) can be neglected with respect 
to any characteristic qp energy. In fact, as seen from the frequency scale in Fig. 1, 
values of order $\Omega\simeq 0.01-0.1$ ps$^{-1}$ should be expected to be consistent with
such a picture, and this assumption will be quantitatively tested later on.

Now it is instructive to plot the integrand of Eq. (\ref{16}) for different temperatures.
In Fig. 2 the
panel (a) shows the low temperature behavior, i.e. only the phonon part of the dispersion
range is appreciable; while for $T=0.45$ K (panel (b)) apart from the low temperature phonon
peak, there is a couple of minor peaks around the roton minimum ($\sim$1.93 $\rm\AA^{-1}$)
associated to transitions involving $R^\pm$ rotons. There is also a pretty
small fourth peak centered
around 0.5 $\rm\AA^{-1}$, which corresponds to phonon$\leftrightarrow$roton transitions.
For $T=$0.6 K (panel (c)) we observe that both phonon peaks turn out to be of the same order
and much smaller than the roton peaks. Finally, for temperatures above 0.9 K only the phonon
peak related to transitions is visible, remaining much smaller than the couple of roton 
peaks (panel (d)).

The above description shows that 
for temperatures up to 0.4 K, only the phonon part of the dispersion curve is significant
in evaluating the expression (\ref{16}), consequently if we set $q_1=k$ and 
$\omega_k=c_sk$ in Eq. (\ref{16}), we  obtain the result (\ref{12}) as expected.
On the other hand, for temperatures above 0.6 K, only the portion of the dispersion curve
around the roton minimum makes a significant contribution to the integrand in Eq. (\ref{16}),
then one can safely make use of the usual approximations in roton calculations.\cite{lan}
In fact, the Landau parabolic approximation, $\omega_k=\Delta/\hbar+
\hbar(k-k_0)^2/2\mu$, leads to two values for $q_j$, $q_1=k$ and
$q_2=2k_0-k$; in addition the Bose distribution in Eq. (\ref{16}) can be well approximated by
the Boltzmann distribution, and the range of integration can be extended in such an equation
from $-\infty$ to $+\infty$ with negligible error. Under such approximations we have
 in the roton dominated regime,
\begin{equation}
D_0\simeq\frac{38}{105}\frac{\hbar^3k_0^5}{m^2c_s^2}\,\,e^{-\Delta/k_BT}.
\label{18}
\end{equation}
Rayfield and Reif\cite{reif}
found a similar expression for the temperature dependence of the longitudinal friction
coefficient in agreement with their experimental data on large vortex rings
obtained for temperatures up to 0.67
K. The validity of such an expression was later\cite{dijk} extended to temperatures nearly
below 1.3 K. The only difference 
between our result (\ref{18}) and the Rayfield-Reif formula
 stems from the coefficient of the exponential $e^{-\Delta/k_BT}$, which
in their case is about 7 \% lower than ours. Such a difference, however, turns out to
be of the order of the experimental uncertainties. 
Another common formula in the literature\cite{h-v,bar} is $D=\rho_n v_G\sigma_\parallel$,
where $v_G=
\langle|\partial\omega_q/\partial q|\rangle=\sqrt{2k_BT/\pi\mu}$ is the average group
velocity of thermal rotons, and $\sigma_\parallel$ is a longitudinal scattering length which
can be approximated by the constant value 8.38 $\rm\AA$ for temperatures below 1.3 K.
Taking into account the roton approximation for $\rho_n$, such a formula practically yields  
the same friction coefficient values as those given by the Rayfield-Reif expression.

Fig. 3 corresponds to plots of the friction coefficient $D$ versus temperature in the
range 0.4 K-1.3 K. The circle points were calculated as the sum of
 the Rayfield-Reif expression plus the
Iordanskii expression, Eq. (\ref{12}); 
note that above 0.6 K the former clearly dominates giving
practically a straight line.
The solid line corresponds to $D_0$ (Eq. (\ref{16})), while 
the values of
 $D_\Omega$ with $\Omega=$ 0.1 ps$^{-1}$ (Eqs. (\ref{D0}) and (\ref{15})) are given by
the dash-dotted line, and the corresponding values for $\Omega=$ 0.01 ps$^{-1}$
were not plotted since they lie in
between the solid line and the center of the circle points
 in the whole temperature range.
From Fig. 3 we may also realize  that the expression (\ref{18}) gives a good approximation
to Eq. (\ref{16}) for temperatures above 0.6 K, where the solid line practically becomes a 
straight line. But, it is clear that Eq. (\ref{18}) does not take into account any phonon
process or, in other words, we may conclude that the contribution of the phonon peaks
in Fig. 2(c-d) must not have an appreciable weight in 
the integrand of Eq. (\ref{16}). On the other hand,
we have found that the weight of roton $R^+\leftrightarrow R^-$ transitions is important.\cite
{sam} In fact, the dashed line in Fig. 3 corresponds to the values extracted from
Eq. (\ref{16}) considering only $q_1=k$, that is, 
 neglecting both, phonon-roton and roton $R^+\leftrightarrow R^-$
transitions. 
Notice that such a curve shows for $T>0.6$ K the same temperature dependence,
$e^{-\Delta/k_BT}$, (i.e., the same slope in Fig. 3) as the expression (\ref{18}), 
but the 
coefficient accompanying the exponential turns out to be substantially minor ($\sim$50\%).

For temperatures above 1.3 K, the value of the friction coefficient is deduced from
experiments involving attenuation of second sound.\cite{bar} Fig. 4 shows that the microscopic
and the phenomenological coefficients become increasingly differentiated above 1.3 K, with
important uncertainties for higher temperatures, particularly in the case of the microscopic
coefficient. Again we can see that the values for $D_0$ (solid line) turn out to be slightly
greater than those for $D_\Omega$ with $\Omega=$ 0.01 ps$^{-1}$ (dash-double-dotted line), 
and both keep in excellent agreement with the microscopic values up to temperatures
about 1.5 K. The discrepancy observed for higher temperatures could be ascribed to
a temperature breakdown of the non-interacting qp gas picture, since the higher experimental
values of vortex friction are consistent with an emergence of interaction effects in the qp
gas.
In addition, possible
effects of  disregarded dragging of normal fluid by the vortex, and temperature dependence 
of $\Omega$, could be affecting our results as well. 
The effect of neglecting vortex-induced qp transitions is again important, as seen from
the dashed line which turns out to be about 50 \% lower than $D_0$. 
Most of such transitions are of course between $R^\pm$ rotons since, 
if we neglect from $D_0$ in Eq. (\ref{16})
only phonon-roton transitions, we obtain 
that the solid curve in Fig. 4 would be decreased less than 8 \%.

\section{Summary and conclusions}
\label{sec4}
We have studied the longitudinal friction force acting on a vortex line in superfluid
$^4$He. Our treatment was based on the analogy between such a vortex dynamics and that of
the quantal Brownian motion  of a charged point particle in a uniform magnetic field.
The friction arising from the scattering of superfluid qp excitations by the vortex was 
taken into account in the form of a translationally invariant interaction potential which,
expanded to first order in the vortex velocity operator, gives rise to vortex transitions
between nearest Landau levels. The corresponding friction coefficient was shown to be, in the
limit of elastic scattering
of qp excitations by the vortex, equivalent to that arising from the 
Iordanskii formula. To evaluate such a coefficient beyond the phonon dominated regime, we 
first noticed the singularities in the integrand defining the friction coefficient stemming 
from the zeros of the qp group velocity (roton and maxon points). Then, from the observation
that such divergences could only be avoided given a suitable functional form for
 the vortex-qp scattering amplitude, a very
simple one-parameter form was proposed for the whole qp dispersion
range. The unknown scale parameter was adjusted to yield exactly the
Iordanskii result for the friction coefficient in the phonon dominated regime, and thus
such a coefficient was finally evaluated for higher temperatures
taking into account the whole qp dispersion curve.

We have analyzed different possibilities for the calculation of the friction coefficient
beyond the phonon range, namely,

\noindent (i) Our expression $D_0$, which assumes elastic scattering of qp excitations by the
vortex ($\Omega\rightarrow 0$) and yields an excellent agreement with the experimental values
up to temperatures about 1.5 K.

\noindent (ii) Our expression $D_\Omega$ with cyclotron frequency values extracted from
recent theories ($\Omega$ ranging between 0.01 and 0.1 ps$^{-1}$), the lower bound of 
$\Omega$ yielding practically the same results as (i) and the upper bound underestimating
the experimental data about 20 \%.

We have also estimated, in the roton dominated regime,
the relative importance of vortex-induced qp transitions on the friction process.
We have found that the qp-vortex interaction processes with no change in qp 
momentum magnitude, are responsible for about 50 \% of the value of the friction coefficient,
 and the remaining contribution arises from roton 
$R^+\leftrightarrow R^-$ transitions ($\sim$42 \%) and roton-phonon transitions ($\sim$8 \%).

In conclusion, we have presented a rather simple theoretical
model for  the longitudinal friction on 
vortices in helium II, which yields excellent agreement with the values derived from
 experimental data
and provides important insights about the microscopic mechanisms of mutual friction.

\begin{figure}
        \begin{center}
\epsfig{file=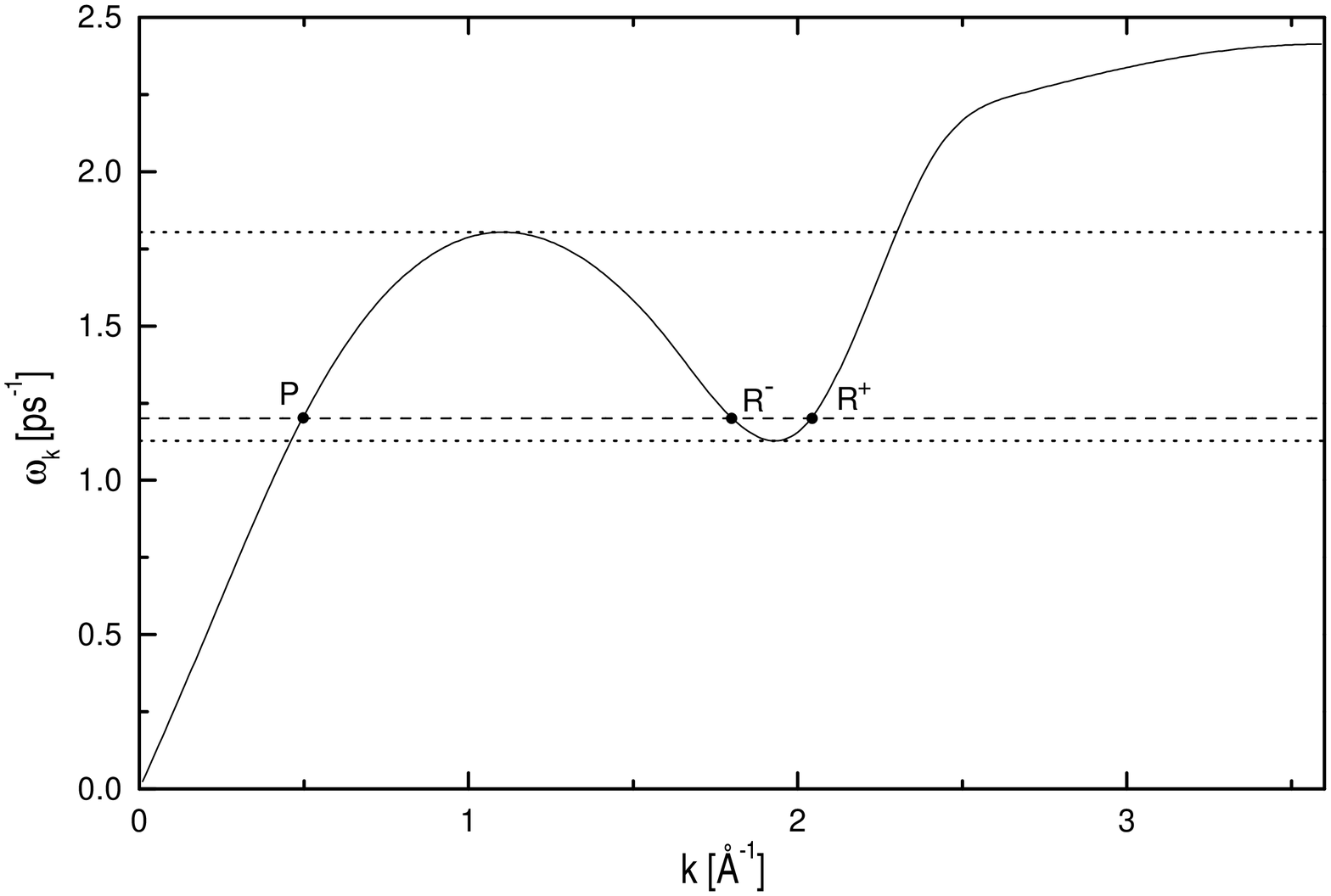,width=\columnwidth}
\caption{\label{fig1}Dispersion curve for elementary excitations in helium II (from
Ref. 22). The  dotted lines separate the regions where the equation
$\omega_q=\omega_k$ changes its number of solutions. The points $P$, $R^-$ and $R^+$ on the
dashed line illustrate about possible vortex-induced qp transitions.}
\end{center}
\end{figure}

\begin{figure}
\centering
\epsfig{file=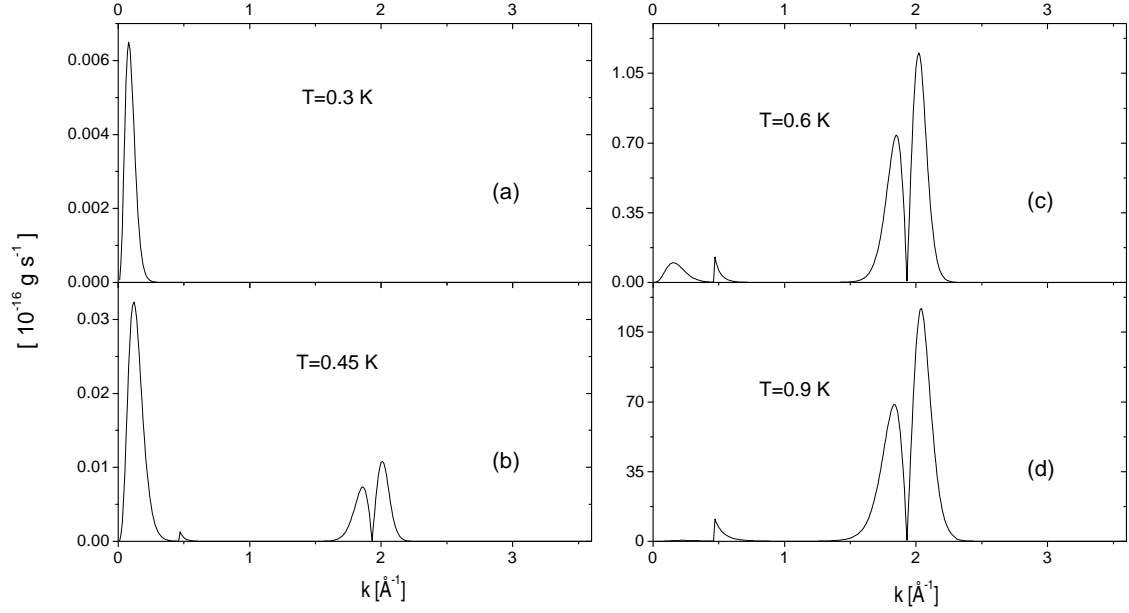,width=\columnwidth}
\caption{\label{fig2} Integrand of $D_0$ (Eq. (\ref{16})) for different temperatures. 
Qualitatively similar plots are obtained for the integrand of $D_\Omega$
with $\Omega$ of order $0.01-0.1$ ps$^{-1}$.}
\end{figure}

\begin{figure}
        \centering
\epsfig{file=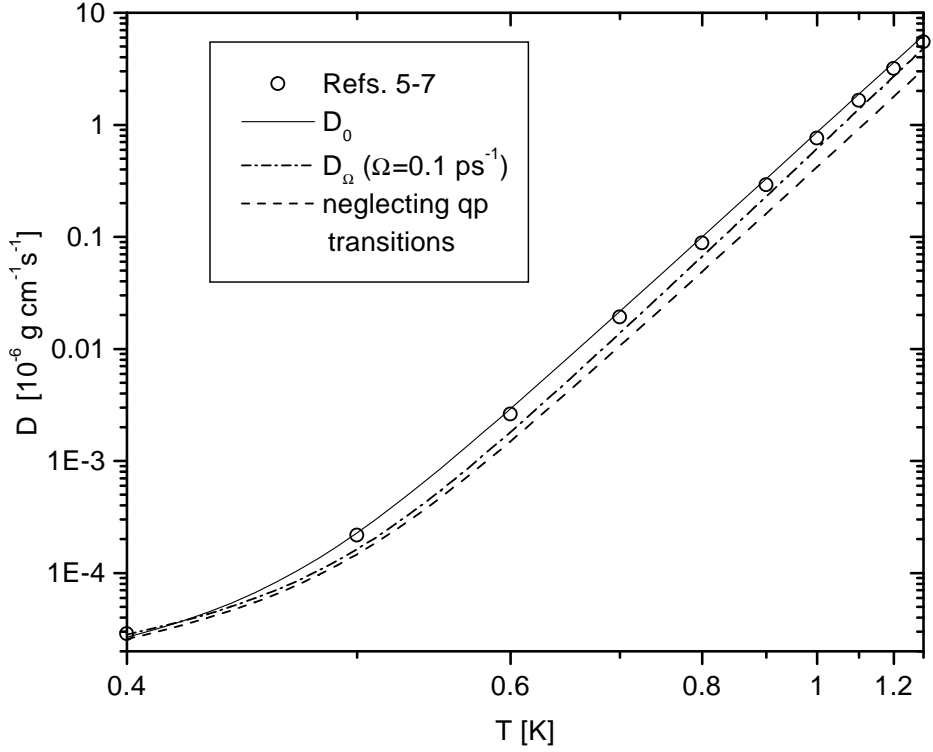,width=\columnwidth}
\caption{\label{fig3} Friction coefficient $D$ versus temperature. The horizontal 
and vertical scales are respectively $-T^{-1}$ and $\log_{10}D$. The circle points
corresponds to the Rayfield-Reif roton expression (Ref. 5, Eq. (28)) plus
the Iordanskii phonon expression (\ref{12}). The solid and dash-dotted
lines correspond respectively to $D_0$ given by Eq.
(\ref{16}) and to 
$D_\Omega$ with $\Omega=$ 0.1 ps$^{-1}$ (Eqs. (\ref{D0}) and (\ref{15})).
The dashed line corresponds to Eq. (\ref{16}) without phonon-roton and
roton $R^+\leftrightarrow R^-$ transitions.}
\end{figure}

\begin{figure}
        \centering
\epsfig{file=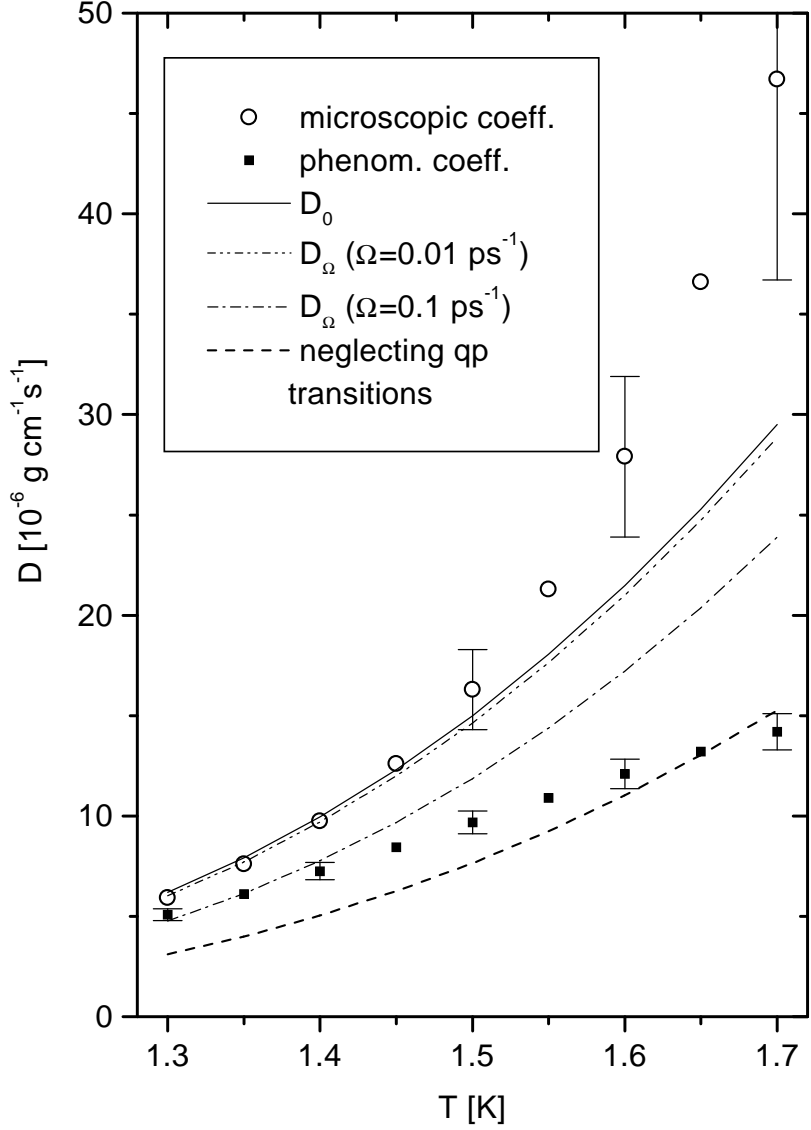,height=18.6cm}
\caption{\label{fig4} Friction coefficient $D$ versus temperature.
The circle points correspond to the values of the microscopic coefficient given in table III
of Ref. 3, whereas the square points correspond to the
values of the phenomenological coefficient
given in table II of the same paper. The dash-double-dotted 
line corresponds to $D_\Omega$ with $\Omega=$ 0.01 ps$^{-1}$ (Eqs. (\ref{D0}) and (\ref{15})),
while the remaining lines are the same as in Fig. 3.
}

\end{figure}

\begin{references}
\bibitem{he3} The effect of the $^3$He impurities on vortex friction, which yields the
most important contribution for the lowest temperature range,\cite{reif} is not taken
into account in our treatment.
\bibitem{not1} Rotons at the left side of the roton minimum have their group velocity
antiparallel to their momentum and are called $R^-$ rotons; those at the right side of
the roton minimum have their group velocity parallel to their momentum and are called
$R^+$ rotons.
\end{references}
\end{document}